\documentclass{article}
\usepackage{graphicx}
\usepackage{amsmath}
\usepackage{amsfonts}
\usepackage{amssymb}

\begin{document}

\title{Environment-independent decoherence rate in classically chaotic systems.}
\author{Rodolfo A. Jalabert$^{1}$ \ and Horacio\ M. Pastawski$^{1,2}$\\$^{1}$\textit{Institut de Physique et Chimie des Mat\'{e}riaux de Strasbourg, }\\\textit{UMR 7504, CNRS-ULP, 23 rue du Loess, 67037 Strasbourg Cedex, France}\\$^{2}$\textit{Facultad de Matem\'{a}tica, Astronom\'{\i}a y F\'{\i}sica,
Universidad }\\\textit{Nacional de C\'{o}rdoba, Ciudad Universitaria, 5000 C\'{o}rdoba, Argentina}}
\maketitle
\begin{abstract}
We study the decoherence of a one-particle system, whose classical
correpondent is chaotic, when it evolves coupled to a weak quenched
environment. This is done by analytical evaluation of the Loschmidt Echo,
(i.e. the revival of a localized density excitation upon reversal of its time
evolution), in presence of \ the perturbation. We predict an \ exponential
decay for the Loschmidt Echo with a (decoherence) rate which is asymptotically
given by the mean Lyapunov exponent of the classical system, and \ therefore
independent of the perturbation strength, within a given range of strengths.
Our results are consistent with recent experiments of Polarization Echoes in
nuclear magnetic resonance and preliminary numerical simulations. PACs:
03.65.Sq, 05.45.+b, 05.45.Mt, 03.67.-a
\end{abstract}

The coupling of a system to environmental degrees of freedom plays an
important role in many areas of physics. Already on a classical level, it
leads to fluctuations, damping and irreversibility. In quantum mechanics the
environmental coupling induces decoherence, destroying quantum superpositions
and reducing pure states to a mixture of states \cite{dissipation}. It is then
not surprising that the concepts of environment, decoherence and
irreversibility have been the object of scholar discussions for long time
\cite{HPZ}. Renewed interest has been fostered by the crucial role that
decoherence plays in the problem of quantum computation \cite{divin95} and by
the technical advances that make it possible to perform experiments envisioned
as \emph{gedanken }for long time.

Experiments with Rydberg atoms in a microwave cavity \cite{brune96a} allow to
observe the progressive decoherence in a quantum measurement problem, while
analysis of conductances through semiconductor microstructures \cite{bshmu}
make it possible to address the ``which path'' problem in a solid state
environment. In addition, nuclear magnetic resonance (NMR) offers unlimited
possibilities for the study of decoherence and irreversibility in a tailored
environment. The phenomenon of \textit{spin echo}, shows how an
\textit{individual spin}, in an ensemble, loses its ``phase memory''
\cite{Hahn} as a consequence of the interaction with other spins that act as
an ``environment''. The failure of recovering the initial ordered state in a
time scale $T_{2},$ manifest the effect of a complex (many-spin) environment
on the reversibility of simple systems. A further conceptual breakthrough was
enabled by experiments that revert\cite{ZME} and control\cite{Exper-Chaos} the
\textit{whole} entangled\textit{ state }of the strongly interacting nuclear
spins to obtain the NMR \textit{Polarization Echo.} We will briefly discuss
the physics involved.

A local spin excitation $\left|  \psi\right\rangle $ created at time $t\!=\!0$
spreads out through the crystal under the action of a \emph{many-spin}
Hamiltonian $\mathcal{H}_{0}$ \ allowing exchange between spins. This complex
quantum evolution is macroscopically assimilated to a
``spin-diffusion''\cite{spin-dif} process (consistently with the usual
hypothesis of microscopic chaos describing many particle systems
\cite{Gaspard-NATURE}). At time $t,\!$ a radio-frequency pulse sequence
produces a new effective Hamiltonian $-(\mathcal{H}_{0}+\Sigma)$. Here
$\Sigma,$ a perturbation containing the pulse imperfections and residual
interactions with additional spins, can be made very small\cite{Exper-Chaos}.
Hence, after the pulse at $t,$ one gets an implementation of the
\emph{gedanken }backwards dynamics proposed by Loschmidt in his argument
against the Boltzmann's H-theorem. At time\thinspace\thinspace$2t,$ one
measures a maximum in the return probability to the initial state that we call
Loschmidt Echo (LE),
\begin{equation}
M(t)=\left|  \left\langle \psi\right|  \operatorname{e}^{\mathrm{i}\left(
\mathcal{H}_{0}+\Sigma\right)  t/\hbar}\operatorname{e}^{-\mathrm{i}%
\mathcal{H}_{0}t/\hbar}\left|  \psi\right\rangle \right|  ^{2}.\label{eq:M_LE}%
\end{equation}
The build up of the LE depends on a precise interference between the
``diffusive'' wave-packets $\operatorname{e}^{-\mathrm{i}\mathcal{H}%
_{0}t/\hbar}\left|  \psi\right\rangle $ and $\operatorname{e}^{-\mathrm{i}%
\left(  \mathcal{H}_{0}+\Sigma\right)  t/\hbar}\left|  \psi\right\rangle ,$
that is degraded by $\Sigma$. Clearly, $M(t)$ should be a decreasing function
of the elapsed time $t$ before the reversal of $\mathcal{H}_{0}$ with a
decoherence rate $1/\tau_{\phi}<1/T_{2}$. A surprising outcome of the
experiment\cite{Exper-Chaos} is that, for small $\Sigma$'s, $1/\tau_{\phi}$
\emph{only depends on the intrinsic properties of the system} (that is, on
$\mathcal{H}_{0}$).

In this work, we develop a simple analytical model exhibiting the independence
of the decoherence rate on the perturbation found in the experiment. The
system is represented by a \emph{single-particle Hamiltonian} $\mathcal{H}%
_{0}$ whose underlying classical dynamics is strongly chaotic. This is clearly
an oversimplification respect to the many-body Hamiltonian of interest, but
still it introduces enough complexity in the intrinsic system (quantified by
its mean Lyapunov exponent $\lambda$) which is absent in simpler dissipative
systems, where $\mathcal{H}_{0}$ is integrable. Placing ourselves between the
limits of a trivial and a many-body $\mathcal{H}_{0}$ allows us to have a
tractable model and explore the influence of classical chaos in quantum
dynamics. To account for ``non-inverted'' part of the Hamiltonian evolution we
consider an Hermitian operator $\Sigma$ representing the coupling with a
quenched environment acting in the backward evolution (from $t$ to $2t$). This
approach is not only consistent with the experimental situation but it is also
able to provide a new insight into the problem of decoherence because the
calculation can be handled within the precise framework of the Schr\"{o}dinger
equation. In contrast, most of the previous studies of decoherence use
extremely simple Hamiltonian systems \cite{dissipation} interacting with a
dissipative environment (e.g. stochastic noise) which justifies the use of a
master equation for the reduced density matrix. In this context, by discussing
the entropy growth of a dissipative system, Zurek and Paz \cite{ZP} hinted at
the importance of the chaotic classical dynamics in setting the characteristic
time scales for decoherence. This is consistent, under conditions that we
specify bellow, with our results for the time decay of $M(t).$ However, since
we use a purely Hamiltonian approach, our conceptual framework is very different.

As in the experiment, we start with a localized state in a $d$-dimensional space$,$%

\begin{equation}
\psi(\mathbf{\bar{r}};t=0)=\left(  \tfrac{1}{\pi\sigma^{2}}\right)  ^{d/4}%
\exp\left[  \mathrm{i}\mathbf{p}_{0}\cdot\left(  \mathbf{\bar{r}}%
-\mathbf{r}_{0}\right)  -\tfrac{1}{2\sigma^{2}}\left(  \mathbf{\bar{r}%
}-\mathbf{r}_{0}\right)  ^{2}\right]  , \label{eq:initialwf}%
\end{equation}
centered at $\mathbf{r}_{0}$, with dispersion $\sigma.$ The momentum
$\mathbf{p}_{0}$ selects the energy range of the excitation. This choice also
renders the calculations tractable. The time evolution of the initial state is
best described using the propagator $K(\mathbf{r},\mathbf{\bar{r}%
};t)=\left\langle \mathbf{r}\right|  \operatorname{e}^{-\mathrm{i}%
\mathcal{H}t/\hbar}\,\left|  \mathbf{\bar{r}}\right\rangle $ by%

\begin{equation}
\psi(\mathbf{r};t)=\int\mathrm{d}\mathbf{\bar{r}}\ K(\mathbf{r},\mathbf{\bar
{r}};t)\ \psi(\mathbf{\bar{r}};0)\ . \label{eq:defprop}%
\end{equation}
Using the Hamiltonian $\mathcal{H}_{0}+\Sigma$ or $\mathcal{H}_{0}$ in the
propagator $K$ yields $\psi_{\mathcal{H}_{0}+\Sigma}$ or $\psi_{\mathcal{H}%
_{0}}$, respectively. We take $\Sigma$ as a static disordered potential given
by $N_{i}$ impurities with a Gaussian potential characterized by the
correlation length $\xi$,%

\begin{equation}
\Sigma=\tilde{V}(\mathbf{r})=\sum_{\alpha=1}^{N_{i}}\frac{u_{\alpha}}{(2\pi
\xi^{2})^{d/2}}\ \exp\left[  -\tfrac{1}{2\xi^{2}}(\mathbf{r}\!-\!\mathbf{R}%
_{\alpha})^{2}\right]  . \label{eq:irrevpart}%
\end{equation}
The independent impurities are uniformly distributed with density $n_{i}%
=N_{i}/\mathtt{V}$, ($\mathtt{V}$ is the sample volume). The strengths
$u_{\alpha}$ obey $\langle u_{\alpha}u_{\beta}\rangle=u^{2}\delta_{\alpha
\beta}$. This assumptions simplifies analytical evaluation of the ensemble
average of the observable $M(t)$. We stress that we are not simply describing
the physics of disordered systems (which is obviously phase coherent), since
the potential $\tilde{V}(\mathbf{r})$ acts in the backwards propagation but
not in the forward path.

As a calculational tool, we use the semiclassical approximation for
$K(\mathbf{r},\mathbf{\bar{r}};t)$, as the sum over all the classical
trajectories $s(\mathbf{r},\mathbf{\bar{r};}t)$ joining the points
$\mathbf{\bar{r}}$ and $\mathbf{r}$ in a time $t$ \cite{LesHou89, Gutz-book};%
\begin{align}
K(\mathbf{r},\mathbf{\bar{r}};t)  &  =\sum_{s(\mathbf{r},\mathbf{\bar{r}}%
;t)}K_{s}(\mathbf{r},\mathbf{\bar{r}};t)\ ,\,\,\,\,\,\mathrm{with}%
\label{eq:allsemi}\\
K_{s}(\mathbf{r},\mathbf{\bar{r}};t)  &  =\left(  \tfrac{1}{2\pi
\mathrm{i}\hbar}\right)  ^{d/2}C_{s}^{1/2}\exp{\left[  \tfrac{\mathrm{i}%
}{\hbar}\mathcal{S}_{s}(\mathbf{r},\mathbf{\bar{r}};t)-\tfrac{\mathrm{i}\pi
}{2}\mu_{s}\right]  },\nonumber
\end{align}
valid in the limit of large energies for which the de Broglie wave-length
($\lambda_{F}=2\pi/k=2\pi\hbar/p_{0}$) is the minimal length scale.
$\mathcal{S}$ is the Hamilton principal function (action) specified by the
integral of the Lagrangian $\mathcal{S}_{s}(\mathbf{r},\mathbf{\bar{r}%
};t)=\int_{0}^{t}$ $\mathrm{d}\bar{t}\mathcal{L}$ along the classical path,
$\mu$ is the Maslov index that counts the number of conjugate points (focal
points), and the Jacobian $C_{s}=\left|  \det\mathcal{B}_{s}\right|  $
accounting for the conservation of the classical probability, is expressed in
terms of the initial and final position components $j$ and $i$ as
$(\mathcal{B}_{s})_{ij}=-\partial^{2}\mathcal{S}_{s}/\partial r_{i}%
\partial\bar{r}_{j}$. This approximation gives the wave function with great
accuracy up to very long times\cite{sc-dynamics}. Besides, \ it provides the
leading-order corrections in $\hbar$ due to $\tilde{V}(\mathbf{r})$, in the
limit of $k\xi\gg1,$ from the classical perturbation theory for the
actions\cite{JMP,Kbook}.

Using the initial wave-function of Eq. (\ref{eq:initialwf}) and Eq.
(\ref{eq:allsemi}) we can write,%

\begin{equation}
\psi(\mathbf{r};t)=(4\pi\sigma^{2})^{d/4}\sum_{s(\mathbf{r,r}_{0};t)}%
K_{s}(\mathbf{r},\mathbf{r}_{0};t)\ \exp{\left[  -\tfrac{\sigma^{2}}%
{2\hbar^{2}}\left(  \mathbf{\bar{p}}_{s}-\mathbf{p}_{0}\right)  ^{2}\right]
}, \label{eq:psisemi}%
\end{equation}
where we used that $\left.  \partial\mathcal{S}/\partial\bar{r}_{i}\right|
_{\mathbf{\bar{r}=r}_{0}}=-\bar{p}_{i}$, and neglected the second order terms
of $\mathcal{S}$ in $(\mathbf{\bar{r}}-\mathbf{r}_{0})$. This is justified
under the assumption that $\xi\gg\sigma\gg\lambda_{F},$ i.e. an initial
wave-packet concentrated in a smaller scale than the fluctuations of
$\tilde{V}(\mathbf{r})$. In Eq.~(\ref{eq:psisemi}), only trajectories with
initial momentum $\mathbf{\bar{p}}_{s}$ closer than $\hbar/\sigma$ to
$\mathbf{p}_{0}$ are relevant for the propagation of the wave-packet.

The semiclassical approximation to the LE is:%
\begin{align}
M(t)  &  =\left(  \tfrac{\sigma^{2}}{\pi\hbar^{2}}\right)  ^{d}\ \ \left|
\int\mathrm{d}\mathbf{r}\sum_{s,\tilde{s}}C_{s}^{1/2}C_{\tilde{s}}^{1/2}%
\exp{\left[  \frac{\mathrm{i}}{\hbar}\left(  \mathcal{S}_{s}-\mathcal{S}%
_{\tilde{s}}\right)  -\frac{\mathrm{i}\pi}{2}\left(  \mu_{s}-\mu_{\tilde{s}%
}\right)  \right]  }\right. \nonumber\\
&  \ \left.  \exp{\left[  -\tfrac{\sigma^{2}}{2\hbar^{2}}\left(  \left(
\mathbf{\bar{p}}_{s}-\mathbf{p}_{0}\right)  ^{2}+\left(  \mathbf{\bar{p}%
}_{\tilde{s}}-\mathbf{p}_{0}\right)  ^{2}\right)  \right]  \,\,}\right|  ^{2},
\label{eq:oversqsemi}%
\end{align}
and involves two spatial integrations and four trajectories.

The perfect echo of $\Sigma=0$ is already obtained considering only
trajectories $s=\tilde{s}$ \ which leaves aside terms with a highly
oscillating phase:
\begin{equation}
M_{\Sigma=0}(t)=\left(  \tfrac{\sigma^{2}}{\pi\hbar^{2}}\right)  ^{d}\ \left|
\ \int\mathrm{d}\mathbf{r}\sum_{s}C_{s}\exp{\left[  -\tfrac{\sigma^{2}}%
{\hbar^{2}}\left(  \mathbf{\bar{p}}_{s}-\mathbf{p}_{0}\right)  ^{2}\right]
}\right|  ^{2}=1. \label{eq:mwse0}%
\end{equation}
The integration requires the change from final position variable $\mathbf{r}$
to initial momentum $\mathbf{\bar{p}}$ using the Jacobian $C$.

In the coupled case the square modulus requires a the second integration
variable $\mathbf{r}{^{\prime}.}$ We see that only the terms with slightly
perturbed trajectories $s=\tilde{s}$ (as well as $s^{\prime}=\tilde{s}%
^{\prime}$) survive the average over impurities. Thus,%
\begin{align}
M(t)  &  \simeq\left(  \tfrac{\sigma^{2}}{\pi\hbar^{2}}\right)  ^{d}%
\ \int\mathrm{d}\mathbf{r}^{{}}\int\mathrm{d}\mathbf{r}^{\prime}\ \sum
_{s,{s}^{\prime}}C_{s}C_{{s}^{\prime}}\exp{\left[  \frac{\mathrm{i}}{\hbar
}\left(  \Delta\mathcal{S}_{s}-\Delta\mathcal{S}_{s^{\prime}}\right)  \right]
}\label{eq:oversqdi}\\
&  \exp{\left[  -\tfrac{\sigma^{2}}{\hbar^{2}}\left(  \left(  \mathbf{\bar{p}%
}_{s}-\mathbf{p}_{0}\right)  ^{2}+\left(  \mathbf{\bar{p}}_{s^{\prime}%
}-\mathbf{p}_{0}\right)  ^{2}\right)  \right]  }\ ,\nonumber
\end{align}
where $\Delta\mathcal{S}_{s}=-\int_{0}^{t}\mathrm{d}\bar{t}\ \tilde
{V}(\mathbf{q}_{s}(\bar{t}))$ and $\Delta\mathcal{S}_{s^{\prime}}$ are the
phase differences, along the trajectories $s$ and $s^{\prime},$ resulting from
the perturbation $\tilde{V}$. From Eq.~(\ref{eq:oversqdi}) we see that we can
decompose $M$ into
\begin{equation}
M(t)=M^{\mathrm{nd}}(t)+M^{\mathrm{d}}(t)\ , \label{eq:decomposition}%
\end{equation}
where the first term (non-diagonal) contains trajectories $s$ and ${s}%
^{\prime}$ exploring different regions of phase space, while in the second
(diagonal) ${s}^{\prime}$ remains close to $s$.

In the \emph{non-diagonal} term the impurity average can be done independently
for $s$ and ${s}^{\prime}$. For trajectories longer than $\xi$ the phase
accumulation $\Delta\mathcal{S}_{s}$ results from uncorrelated contributions,
and therefore can be assumed to be Gaussian distributed \cite{JMP,Kbook}. The
disorder contribution involved in Eq.~(\ref{eq:oversqdi}) is then given by
\begin{equation}
\langle\exp{\left[  \frac{\mathrm{i}}{\hbar}\Delta\mathcal{S}_{s}\right]
}\rangle=\exp\left[  -\tfrac{1}{2\hbar^{2}}\ \int_{0}^{t}\mathrm{d}\bar
{t}\mathit{\ }\int_{0}^{t}\mathrm{d}\bar{t}^{\prime}\ C_{\tilde{V}}%
(|q_{s}(\bar{t})-q_{s}(\bar{t}^{\prime})|)\right]  \ , \label{eq:gauss_av2}%
\end{equation}
where the correlation function of the disordered potential is%

\begin{equation}
C_{\tilde{V}}(|\mathbf{q}-\mathbf{q}^{\prime}|)=\langle\tilde{V}%
(\mathbf{q})\tilde{V}(\mathbf{q}^{\prime})\rangle=\tfrac{u^{2}n_{i}}{(4\pi
\xi^{2})^{d/2}}\ \exp\left[  -\tfrac{1}{4\xi^{2}}(\mathbf{q}-\mathbf{q}%
^{\prime})^{2}\right]  . \label{eq:corr}%
\end{equation}
The change of variables $q=v\bar{t}$ and $q^{\prime}=v\bar{t}^{\prime}$ yields
two integrals along the trajectory $s$. Since the length $L_{s}$ of the
trajectory is supposed to be much larger than $\xi$, the integral over
$q-q^{\prime}$ can be taken from $-\infty$ to $+\infty$, while the integral on
$(q+q^{\prime})/2$ gives a factor of $L_{s}$. Assuming that the velocity along
the trajectory remains almost unchanged respect to its initial value
$v_{0}=p_{0}/m=L_{s}/t$ , one gets,%
\begin{align}
M^{\mathrm{nd}}(t)  &  \simeq\left(  \tfrac{\sigma^{2}}{\pi\hbar^{2}}\right)
^{d}\ \left|  \ \int\mathrm{d}\mathbf{r}\sum_{s}C_{s}\exp{\left[
-\tfrac{\sigma^{2}}{\hbar^{2}}\left(  \mathbf{\bar{p}}_{s}-\mathbf{p}%
_{0}\right)  ^{2}\right]  }\ \exp\left[  -\tfrac{L_{s}}{2\tilde{l}}\right]
\right|  ^{2}\nonumber\\
&  \simeq\exp\left[  -tv_{0}/\tilde{l}\right]  .
\end{align}
In analogy with disordered systems \cite{JMP,Kbook} we have defined the
typical length over which the quantum phase is modified by the perturbation
as:
\begin{equation}
\tilde{l}=\hbar^{2}v_{0}^{2}\ \left(  \int\mathrm{d}q\ C(\mathbf{q})\right)
^{-1}=\frac{4\sqrt{\pi}\hbar^{2}v_{0}^{2}\xi}{u^{2}n_{i}}.
\label{eq:mfp_gauss}%
\end{equation}
We then see that $M^{\mathrm{nd}}(t)$ has its time scale determined by
$\Sigma$ (through $\tilde{l}$).

In computing the \emph{\ diagonal} term $M^{\mathrm{d}}(t)$ we use the expansion%

\begin{equation}
\Delta\mathcal{S}_{s}-\Delta\mathcal{S}_{{s}^{\prime}}=\int_{0}^{t}%
\mathrm{d}\bar{t}\ \mathbf{\nabla}\tilde{V}(\mathbf{q}_{s}(\bar{t}%
))\cdot\left(  \mathbf{q}_{s}(\bar{t})-\mathbf{q}_{{s}^{\prime}}(\bar
{t})\right)  , \label{eq:phachange}%
\end{equation}
since the trajectories $s$ and ${s}^{\prime}$ remain close to each other. The
difference between the intermediate points of both trajectories can be
expressed using $\mathcal{B}$ :%

\begin{equation}
\mathbf{q}_{s}(\bar{t})-\mathbf{q}_{s^{\prime}}(\bar{t})=\mathcal{B}^{-1}%
(\bar{t})(\mathbf{\bar{p}}_{s}-\mathbf{\bar{p}}_{s^{\prime}})=\mathcal{B}%
^{-1}(\bar{t})\mathcal{B}(t)\left(  \mathbf{r}-\mathbf{r}{^{\prime}}\right)
.\label{eq:operatorB}%
\end{equation}
In the chaotic case the behavior of $\mathcal{B}^{-1}(\bar{t})$ is dominated
by the largest eigenvalue $e^{\lambda\bar{t}}$. Therefore we make the
simplification $\mathcal{B}^{-1}(\bar{t})\mathcal{B}(t)=\exp\left[
\lambda(\bar{t}-t)\right]  I$, where $I$ is the unit matrix and $\lambda$ the
mean Lyapunov exponent. Here, we use our hypothesis of strong chaos which
excludes marginally stable regions\cite{weak-chaos} with anomalous time
behavior. Assuming a Gaussian distribution for the extra phase of
Eq.~(\ref{eq:phachange})%

\begin{align}
\langle\exp{\left[  \frac{\mathrm{i}}{\hbar}\left(  \Delta\mathcal{S}%
_{s}-\Delta\mathcal{S}_{{s}^{\prime}}\right)  \right]  }\rangle &
=\exp[-\tfrac{1}{2\hbar^{2}}\ \int_{0}^{t}\mathrm{d}\bar{t}\int_{0}%
^{t}\mathrm{d}\bar{t}^{\prime}\exp\left[  \lambda(\bar{t}+\bar{t}^{\prime
}-2t)\right]  \nonumber\\
&  C_{\mathbf{\nabla}\tilde{V}}(|\mathbf{q}_{s}(\bar{t})-\mathbf{q}_{s}%
(\bar{t}^{\prime})|)\ \left(  \mathbf{r}-{\mathbf{r}^{\prime}}\right)  ^{2}].
\end{align}
We are now led to consider the ``force correlator''%
\begin{align}
C_{\mathbf{\nabla}\tilde{V}}(|\mathbf{q}-\mathbf{q}^{\prime}|) &
=\langle\mathbf{\nabla}\tilde{V}(\mathbf{q})\cdot\mathbf{\nabla}\tilde
{V}(\mathbf{q}^{\prime})\rangle\nonumber\\
&  =\tfrac{u^{2}n_{i}}{\left(  4\pi\xi^{2}\right)  ^{d/2}}\left(  \frac{d}%
{2}-\left(  \frac{\mathbf{q}-\mathbf{q}^{\prime}}{2\xi}\right)  ^{2}\right)
\exp{\left[  -\tfrac{1}{4\xi^{2}}(\mathbf{q}-\mathbf{q}^{\prime})^{2}\right]
}.
\end{align}
We change from the variables $\bar{t}$ and $\bar{t}^{\prime}$ to the
coordinates $q$ and $q^{\prime}$ along the trajectory $s$ and use the fact
that $C_{\mathbf{\nabla}\tilde{V}}$ is short-ranged (in the scale of $\xi$) to
write,%
\begin{align}
M^{\mathrm{d}}(t) &  \simeq\left(  \tfrac{\sigma^{2}}{\pi\hbar^{2}}\right)
^{d}\int\mathrm{d}\mathbf{r}\int\mathrm{d}{\mathbf{r}^{\prime}}\ \sum_{s}%
C_{s}^{2}\ \exp{\left[  -\tfrac{2\sigma^{2}}{\hbar^{2}}\left(  \mathbf{\bar
{p}}_{s}-\mathbf{p}_{0}\right)  ^{2}\right]  }\exp{\left[  -\tfrac{A}%
{2\hbar^{2}}\left(  \mathbf{r}-\mathbf{r}{^{\prime}}\right)  ^{2}\right]
}\nonumber\\
&  \simeq\left(  \tfrac{\sigma^{2}}{\pi\hbar^{2}}\right)  ^{d}\ \int
\mathrm{d}\mathbf{r}\ \sum_{s}C_{s}^{2}\ \left(  \tfrac{2\pi\hbar^{2}}%
{A}\right)  ^{d/2}\exp{\left[  -\tfrac{2\sigma^{2}}{\hbar^{2}}\left(
\mathbf{\bar{p}}_{s}-\mathbf{p}_{0}\right)  ^{2}\right]  },
\end{align}
where $A=(d-1)u^{2}n_{i}/(4\lambda v_{0}(4\pi\xi^{2})^{(d-1)/2})$ results from
the $\bar{t}$ and $\bar{t}^{\prime}$ integrations of Eq.~(\ref{eq:gauss_av2})
in the limit $\lambda t\gg1$. The last line comes from Gaussian integration
over $(\mathbf{r}-\mathbf{r}{^{\prime}})$. The factor $C_{s}^{2}$ reduces to
$C_{s}$ when we make the change of variables from $\mathbf{r}$ to $\mathbf{p}%
$. In the long-time limit $C_{s}^{-1}\propto e^{\lambda t}$, while for short
times $C_{s}^{-1}=t/m$. Using a form that interpolates between these two
limits we have%
\begin{align}
M^{\mathrm{d}}(t) &  \simeq\left(  \tfrac{\sigma^{2}}{\pi\hbar^{2}}\right)
^{d}\int\mathrm{d}\mathbf{\bar{p}}\left(  \tfrac{2\pi\hbar^{2}}{A}\right)
^{d/2}\left\{  \frac{m}{t}\exp\left[  -\lambda t\right]  \right\}  \exp
{[-}\tfrac{{2\sigma^{2}}}{{\hbar^{2}}}{\left(  \mathbf{\bar{p}}-\mathbf{p}%
_{0}\right)  ^{2}]}\nonumber\\
&  \simeq\overline{A}\exp\left[  -\lambda t\right]  ,\label{eq:oversqdi3}%
\end{align}
with $\overline{{A}}=m/(A^{d/2}t)$. Since the integral over $\mathbf{\bar{p}}$
is concentrated around $\mathbf{p}_{0},$ the exponent $\lambda$ is considered
constant . The coupling $\Sigma$ appears only in the prefactor (through
$\overline{{A}}$) and therefore its detailed description is not crucial in
discussing the time dependence of $M^{\mathrm{d}}$. The $t$ factor in
$\overline{{A}}$ induces a divergence for small $t$. However, our calculations
are only valid in the limit $\lambda t\gg1$. \ Long times, (of the order
of\ the Ehrenfest time $t_{E}=\lambda^{-1}\ln[ka]$ where $a$ is a length
characterizing $\mathcal{H}_{0}$), are also excluded from our analysis since
we run into the failure of the diagonal approximation.

Our semiclassical approach made it possible to estimate the two contributions
of Eq.~(\ref{eq:decomposition}) to $M(t)$. The non-diagonal component
$M^{\mathrm{nd}}(t)$ is the dominant contribution in the limit of small
$\Sigma$. In particular, it makes $M_{\Sigma=0}(t)=1$ in Eq.~(\ref{eq:mwse0}).
The small values of $\Sigma$ are not properly treated in the semiclassical
calculation of the diagonal term $M^{\mathrm{d}}(t)$. While increasing the
coupling $\Sigma$ the crossover from $M^{\mathrm{nd}}$ to $M^{\mathrm{d}}$ is
achieved when $\tilde{l}$ becomes smaller than $v_{0}/\lambda$. This condition
is compatible with the assumption that, in the limit $k\xi\gg1,$ classical
trajectories shorter than the perturbation's ``transport mean-free-path''
$\tilde{l}_{\mathrm{tr}.}=4(k\xi)^{2}\tilde{l}$ are not affected
\cite{JMP,Kbook} by the quenched disorder. For strong $\Sigma$ the
perturbative treatment of the actions is also expected to break down.

We can now establish our main conclusion. In a system that classically
exhibits strong chaos and can be characterized by a mean Lyapunov exponent
$\lambda,$ a small random static perturbation may destroy our control of the
quantum phase at a rate,%

\begin{equation}
\frac{1}{\tau_{\phi}}=-\lim_{t\rightarrow\infty}\frac{1}{t}\ \ln
M(t)=\lambda\ , \label{eq:conclusion}%
\end{equation}
provided that the time is taken in the interval $\lambda^{-1}\ll t\ll t_{E}$,
the perturbation presents long-range potential fluctuations ($k\xi\gg1$) and a
strength quantically strong ($\tilde{l}\ll v_{0}/\lambda$) but classically
weak ($v_{0}/\lambda$ $\ll\tilde{l}_{\mathrm{tr}}$). Notice that the
thermodynamic limit, ${\mathtt{V}\rightarrow\infty},$ is required to take $t$
arbitrarily large.

The various restrictions for the validity of our result provide stringent
conditions for its numerical verification. A disordered system represented by
a tight-binding model with the topology of a torus exhibits a characteristic
time of decay of $M(t)$ depending on the disorder ($\mathcal{H}_{0}$), but not
on $\Sigma$ (giving the change of the magnetic flux piercing the torus)
\cite{CuPaJa}. Though subject to finite size limitations, the results show an
environment independent behavior when the perturbation exceeds a critical
value defined by the parametric correlations of the spectra \cite{Aaron}. Even
if our calculations and these\ preliminary numerical studies deal with
single-particle Hamiltonian, we expect that the generic behavior that we found
is robust when considering Hamiltonians with larger complexity, like the
many-particle case which is most relevant to the experiments motivating our work.

The field of Quantum Chaos deals with signatures of the classical chaos on
quantum properties. The most widely studied properties have been the spectral
correlations\cite{Bohigas}, the wave function scars \cite{scars}, and the
parametric correlations \cite{Aaron}. In contrast, the studies of the temporal
domain have been less developed, mainly because of the lack of clear
quantities as those of steady state\cite{Izrailev}. The work of Peres
\cite{Peres} was a partial success in that direction. He distinguished regular
and irregular dynamics on the basis of the asymptotic properties of a
perturbation dependent overlap when \texttt{V} is finite and applied it to
simple systems. While there have been further attempts to define a dynamical
sensitivity to perturbations in other systems and observables \cite{SchCav},
we are not aware of other predictions of a manifestation\ of the classical
Lyapunov exponent in a \emph{Hamiltonian system}, as we did with Eq.
(\ref{eq:conclusion}). On view\ of this result, we think that the issue of
decoherence by quantum evolution in classically chaotic systems, both with
strong chaos and with marginal stability, deserves a more thorough
examination. Studies with other analytical and numerical techniques should
clarify, among other aspects, the effects of different specific perturbations,
the subtle effects of thermodynamic limits, the corrections due to Anderson
localization, and the different temporal laws observed in one-body and
many-body systems. This understanding of dynamical manifestations of chaos in
the quantum world is decisive in the efforts to limit the experimental effects
of decoherence and irreversibility.

The authors have benefited from many fruitful discussions with O. Bohigas, P.
Leboeuf, C. Lewenkopf, J. P. Paz and K. Richter. HMP is affiliated with
CONICET. We also acknowledge the financial support of the French-Argentinian
program ECOS-Sud.

\end{document}